\documentstyle[art12]{article}

\newfont{\AMSL}{msbm10 scaled \magstep 1}

\def\RA{\mbox{\AMSL R}}

\newcommand{\arctg}{\mathop{\rm arctg}\nolimits}
\newcommand{\ch}{\mathop{\rm ch}\nolimits}
\newcommand{\sh}{\mathop{\rm sh}\nolimits}

\newcommand{\mod}{\mathop{\rm mod}\nolimits}

\newtheorem{de}{Definition}[section]
\newtheorem{lem}{Lemma}[section]
\newtheorem{th}{Theorem}[section]

\title{Propagation of infinitely narrow $\delta$-solitons\/}

\author{V.~G.~Danilov\\
\small Moscow Technical University of Communication and Informatics, Russia\\
\small danilov@amath.msk.ru\\
\and
V.~M.~Shelkovich\\
\small St.-Petersburg State Architecture and Civil Engineering University,
Russia\\
\small shelkv@svm.abu.spb.ru\\}

\date{}

\begin{document}
\maketitle

\setcounter{equation}{0}
\begin{sloppypar}

\begin{abstract}
We construct a definition of the weak solution to KdV type equations 
with small dispersion admitting the zero dispersion limit 
for soliton-like solutions.
Using this definition, we obtain a system of equations 
(the limit problem as the dispersion tends to zero) 
that describes the soliton dynamics. 
\end{abstract}

\section{Introduction and basic results}%1
\label{s1}

    1. It is well known that the Korteweg-de Vries (KdV) equation
\begin{equation}
\label{1}
L_{KdV}[u]=u_t + (u^2)_x + \varepsilon ^2u_{xxx} = 0
\end{equation}
has the one-soliton solution
\begin{equation}
\label{2}
u(x,t,\varepsilon ) =
\frac{3v}{2}\ch^{-2}(\frac{\sqrt{v}}{2}(x-vt)/\varepsilon ), \quad x \in
\RA,
\end{equation}
where $v$ is the soliton velocity. The pointwise limit as
$\varepsilon  \to +0$ of solution (\ref{2}) to the KdV equation is the
discontinuous function $\frac{3v}{2}\chi(x-vt)$, where $\chi(\xi)=1$,
if $\xi=0$ and $\chi(\xi)=0$ if $\xi \ne 0$. The weak asymptotics (\ref{2})
as $\varepsilon  \to +0$, up to $O_{{\cal D}'}(\varepsilon ^2)$, becomes
the {\it infinitely narrow $\delta$-soliton}
\begin{equation}
\label{3}
u_{\varepsilon }(x,t) =A\varepsilon \delta(x-vt), \quad \varepsilon \to +0,
\qquad
A=\frac{3v}{2}\int \ch^{-2}(\xi)\,d\xi=6\sqrt{v},
\end{equation}
and $\delta(x)$ is the Dirac delta function.
Here and in what follows $\int$ denotes an improper integral
from $-\infty$ to $+\infty$.
By $O_{{\cal D}'}(\varepsilon ^{\alpha})$ we denote a distribution from
${\cal D}'(\RA)$ such that for any test function $\varphi(x) \in {\cal D}$
$$
\langle O_{{\cal D}'}(\varepsilon ^{\alpha}), \varphi(x)\rangle
=O(\varepsilon ^{\alpha}),
$$
and $O(\varepsilon ^{\alpha})$ is understood in the ordinary sense.

    We stress once more that here all generalized functions (distributions)
are treated as functionals on the space ${\cal D}(\RA_x)$ and these
functionals depend on the other variables as on parameters.

    It follows from (\ref{2}), (1.3) that we have
$u(x,t,\varepsilon )=O_{{\cal D}'}(\varepsilon )$ as $\varepsilon  \to +0$
and $\varepsilon ^2u_{xxx}= O_{{\cal D}'}(\varepsilon ^3)$.
Therefore, the limit expression (\ref{3}) was interpreted by
V.~P.~Maslov and V.~A.~Tsupin~\cite{1}--\cite{2},
V.~P.~Maslov and G.~A.~Omelyanov~\cite{3}--\cite{4}
as an asymptotic up to $O_{{\cal D}'}(\varepsilon ^2)$ {\it generalized
solution\/} of the Hopf equation
\begin{equation}
\label{4}
L_{H}[u]=  u_t + (u^2)_x =  0,
\end{equation}
which is the {\it limit problem} for the KdV equation. In the same works
the corresponding generalized Hugoniot conditions, of the type
of those for the shock wave front, were obtained.

    It is easy to see that in the framework of the above-mentioned
approach, the exact solution (\ref{2}) to KdV equation (\ref{1})
satisfies the Hopf equation (\ref{4}) in the following sense:
\begin{equation}
\label{5}
L_{H}[u]= O_{{\cal D}'}(\varepsilon ^2),
\end{equation}
because the following equality holds:
\begin{equation}
\label{6}
L_{KdV}[u]-L_{H}[u]= O_{{\cal D}'}(\varepsilon ^2).
\end{equation}

    If, instead of expression (\ref{2}) which is the exact solution of the
KdV equation and approximates the weak asymptotics (\ref{3}), we consider
the function
\begin{equation}
\label{7}
{\tilde u}(x,t,\varepsilon )= A\omega\big(\frac{x-vt}{\varepsilon }\big),
\end{equation}
where $\omega(z) \in C^{\infty}(\RA)$ has a compact support or rapidly
decreases as $|z|\to\infty$, $\int\omega(z)\,dz= 1$, then:

    1) expression (\ref{7}) also has asymptotics (\ref{3}) in the sense
of ${\cal D}'$ as $\varepsilon  \to +0$ (see Section~\ref{s2} for details);

    2) substituting (\ref{7}) into the Hopf equation (\ref{4}),
provided that a certain correlation between the constants $v$ and $A$
is true (the generalized Hugoniot condition), we have
$L_H[{\tilde u}]= O_{{\cal D}'}(\varepsilon ^2)$;

    3) and $u(x,t,\varepsilon )-{\tilde u}(x,t,\varepsilon )
= O_{{\cal D}'}(\varepsilon ^2)$.

    Therefore, an asymptotics up to $O_{{\cal D}'}(\varepsilon^2)$,
i.e., an infinitely narrow $\delta$-soliton-type solution of the KdV
equation (or the Hopf equation which is the limit problem of the KdV
equation), can be sought starting not from the exact solution (\ref{2})
of the KdV equation (which is a regularization of the Hopf equation)
but from an ansatz of the form (\ref{7}) substituted {\it directly\/}
into the Hopf equation.

    Generalizing (\ref{3}), one can seek the solution in the form
$$
u^*(x,t,\varepsilon )= u_0(x,t)
+g(t)\omega\big(\frac{x-\phi(t)}{\varepsilon }\big),
$$
by subsituting this singular ansatz into the Hopf equation.

    However, as we shall see in Section~\ref{s3}.1, such an ansatz
results in the solution with  constant amplitude $g= \mbox{const}$ of the
soliton for $u_0\ne\mbox{const}$, which contradicts the well-known
results about the soliton behaviour. Therefore, generalizing formulae
(\ref{2}), (\ref{7}), (\ref{3}), we can attempt to construct an
asymptotic solution to the KdV equation (or the Hopf equation) of the form
\begin{equation}
\label{8}
u^*_{\varepsilon }(x,t)= u_0(x,t)+g(t){\varepsilon }\delta(x-\phi(t))
+ e(x,t){\varepsilon }\theta(x-\phi(t)), \quad \varepsilon  \to +0,
\end{equation}
where $\theta$ is the Heaviside function. A solution in such a form will
be called an {\it infinitely narrow $\delta$-soliton\/}.

    In order to consider the function $u^*_{\varepsilon }(x,t)$
as a solution to a nonlinear equation we need to define the
rules of substitution of this function into the nonlinear term,
that is, to include the functions ${\varepsilon
}\delta(x-\phi(t)$, \ ${\varepsilon }\theta(x-\phi(t))$, where
$\varepsilon  \to +0$, into an algebra with differentiation.

    2. The rules of substitution of singular ansatzs into
nonlinear equations are finally reduced to the definition of
multiplication of distributions from certain classes. Various
approaches to this problem are discussed, for example,
in~\cite{5}. A well-known approach to the problem of
multiplication of distributions related to the name of
J.~Colombeau~\cite{6},~\cite{7} is based on the construction of
an algebra of new generalized functions. The spaces $C^{\infty}$ and $D'$
are embedded into ${\cal G}$, and moreover, the first embedding is an
algebraic isomorphism. Some of the new generalized functions are associated
with distributions but, in the general case, they are not directly related
to the Schwartz distributions.

    Note that H.~A.~Biagioni and M.~Oberguggenberger (\cite{8})
studied the solution to the KdV equation in the form of an
infinitely narrow soliton (\ref{3}) in terms of the theory of
Colombeau new generalized functions.

    Another approach to this problem is associated with the
names of J.~B.~Livchak~\cite{9}, Li Bang-He~\cite{10},
V.~K.~Ivanov~\cite{11}.  The main idea of this approach is that
products of distributions from a certain class are defined as
asymptotic decompositions, in the weak sense, of products of
approximations of these distributions as $\varepsilon  \to +0$,
where $\varepsilon $ is an approximation parameter, and the
coefficients are distributions.

    Development of ideas of this approach in~\cite{12},~\cite{13} and
especially in~\cite{14} has led to the construction of {\it the weak
asymptotics method\/} which enables one to obtain substantial analytical
results for discontinuous solutions to nonlinear equations and, in
particular, to study the dynamics of propagation and interaction of
singularities of various types. In the framework of this approach the
construction of a singular solution to a nonlinear equation is reduced
to the choice of singular generators of the ansatz, that is, distributions
(generalized functions), to defining the rules of substitution
the solution with the chosen structure into the nonlinear
equation, to setting up a system of equations for unknown smooth
functions and to investigation of the system obtained.

    An essential technical progress achieved in these papers, as
compared with the results of~\cite{9} --~\cite{11}, is the
obtaining of complete asymptotic decompositions of products of
approximations of distributions and, on this basis, of
associative and commutative differential algebras of asymptotic
distributions including subspaces of distributions. The elements
of these algebras are weak asymptotics whose coefficients are
distributions. Thus, in~\cite{14} there was constructed an
associative and commutative algebra ${\cal E}$ of asymptotic
distributions (weak asymptotics) generated by the space of
linear combinations of homogeneous or associated homogeneous
distributions. For example, the singular ansatz (\ref{8}) is an
asymptotic distribution belonging to this algebra.

    An asymptotic generalized infinitely narrow
$\delta$-soliton-type solution to the Hopf equation (\ref{2})
with constant amplitude, and the generalized Hugoniot condition
were constructed in the framework of the technique of
substitution of a singular ansatz into ~\cite{12},~\cite{13}.

    The application of the  weak asymptotics method in~\cite{14}
has made it possible to obtain some substantial results about
the structure of singular solutions to quasilinear strictly
hyperbolic systems and, with certain restrictions, to find the
solution of the problem of separating self-similar singularities
formulated in~\cite{15},~\cite{16}.

    3. Thus, in order to substitute the singular ansatz (\ref{8}) into
the KdV equation (\ref{1}) or the Hopf equation (\ref{4}), we first
substitute there a smooth ansatz of the form
\begin{equation}
\label{9}
u^*(x,t,\varepsilon )= u_0(x,t)
+g(t){\varepsilon }\delta(x-\phi(t),\varepsilon )
+ e(x,t){\varepsilon }\theta(x-\phi(t),\varepsilon ), \quad \varepsilon >0,
\end{equation}
where $u_0(x,t)$, \ $g(t)$, \ $e(x,t)$, \ $\phi(t)$ are the desired
smooth functions, and
$$
{\varepsilon }\delta(x,\varepsilon )= \omega(\frac{x}{\varepsilon }),
\qquad
{\varepsilon }\theta(x,\varepsilon )
= {\varepsilon }\omega_0\big(\frac{x}{\varepsilon }\big)
$$
are smooth approximations of the asymptotic distributions
${\varepsilon }\delta(x)$ and ${\varepsilon }\theta(x)$, respectively.

    Here, the function $\omega(z) \in C^{\infty}(\RA)$ either has a
compact support or decreases sufficiently rapidly as $|z|\to \infty$,
for example,
$|\omega(z)| \le C(1+|z|)^{-3}$ and $\int\omega(z)\,dz= 1$;
$\omega_0(z) \in C^{\infty}(\RA)$, \ $\lim_{z \to +\infty}\omega_0(z)= 1$,
\ $\lim_{z \to -\infty}\omega_0(z)= 0$.

    Then we have in the sense of ${\cal D}'(\RA)$ (see the notation above)
$$
\varepsilon \delta(x,\varepsilon )=\varepsilon \delta(x)
+ O_{{\cal D}'}(\varepsilon ^2), \quad
\varepsilon \theta(x,\varepsilon )=\varepsilon \theta(x)
+ O_{{\cal D}'}(\varepsilon ^2), \qquad \varepsilon  \to +0.
$$
For more details, see Section~\ref{s2}.

    Now, by analogy to (\ref{5}), (\ref{6}), we can introduce the
definition of the asymptotic generalized solution of the form (\ref{8}).
Namely, we call {\it asymptotic distribution\/} (\ref{8}) an {\it asymptotic
generalized solution\/} of the Hopf equation (\ref{9}) if its approximation
(\ref{9}) satisfies the relation
\begin{equation}
\label{10}
L_{H}[u^*(x,t,\varepsilon )]= O_{{\cal D}'}(\varepsilon ^2)
\end{equation}
and, respectively, an asymptotic generalized solution to the KdV equation
(\ref{1}), if
\begin{equation}
\label{11}
L_{KdV}[u^*(x,t,\varepsilon )]= O_{{\cal D}'}(\varepsilon ^2),
\end{equation}
which, in fact, is the same, since
$$
L_{KdV}[u^*(x,t,\varepsilon )]-L_{H}[u^*(x,t,\varepsilon )]
= O_{{\cal D}'}(\varepsilon ^2).
$$

    It is easiy to see that  our definition of the solution can
{\it depend\/} on the choice of approximations
$\frac{1}{\varepsilon }\omega(\frac{x-\phi(t)}{\varepsilon })$ and
$\omega_0(\frac{x-\phi(t)}{\varepsilon })$ to the distributions
$\delta(x-\phi(t))$, and $\theta(x-\phi(t))$, respectively.
Actually, as we shall see later, the dynamics of solution of the
type (\ref{8}) {\it is independent\/} of the approximation of
the Heaviside function $\omega_0(\frac{x-\phi(t)}{\varepsilon })$.

    Definitions (\ref{10}) and (\ref{11}) also imply that, in fact,
in order to construct asymptotic generalized solutions satisfying
(\ref{10}) or (\ref{11}), we need not calculate products of generalized
functions (and, in general, any other nonlinearities) to a high accuracy.
Indeed, the substitution of the exact one-soliton solution of the KdV
equation into the Hopf or KdV equations provides just the accuracy
$(O_{{\cal D}'}(\varepsilon ^2))$ corresponsing to~(\ref{10}). On the
other hand, it is clear that we cannot deal with a lesser accuracy, since
the soliton solution contains terms of order $O_{{\cal D}'}(\varepsilon)$.

    In our notation, we can say that, within the framework of algebraic
constructions related to generalized functions, nonlinear expressions are
usually calculated up to $O_{{\cal D}'}(\varepsilon^\infty)$, which is
necessary to define associative and commutative algebras of generalized
functions.

    A distinction of the method of weak asymptotics (which, undoubtedly,
originates from the algebraic constructions mentioned above) is that we
actually deal with approximations. In fact, the difference between the
method of weak asymptotics and the method of ordinary asymptotic expansions
is that the smallness of the remainder is understood in a different way.
Usually, the remainder is assumed to be small in some uniform sense
with sufficient accuracy. Here we assume exactly the same but in the
sense of $O_{{\cal D}'}$.

    In order to obtain the results known from the KdV equation
theory it seems natural to use the function from the formula for
the exact one-soliton solution (\ref{2}) to the KdV equation as
an approximation for $\varepsilon \delta(x-\phi(t),\varepsilon)$.

    The system for the functions $u_0(x,t)$, \ $g(t)$, \ $e(x,t)$,
$\phi(t)$ follows from Definitions (\ref{10}) or (\ref{11}) (this
system will be derived in detail in Section~\ref{3})
\begin{equation}
\label{12}
\begin{array}{rcl}
\displaystyle
u_{0t}+(u^2_0)_x &=&0,\\
\displaystyle
\phi_{t} - 2u_0(\phi(t),t) - \frac{2}{3}g(t) &=&0, \\
\displaystyle
e(\phi(t),t)-\frac{3\sqrt{6}}{2}g_t(t)/g^{3/2}(t) &=&0,\\
\displaystyle
\big(e_{t}(x,t) + 2(u_0(x,t)e(x,t))_x \big)\Big|_{x > \phi(t)} &=&0.
\end{array}
\end{equation}

    It is easy to verify that under the condition $g>0$ (which is an
analog of the admissibility condition in the theory of shock waves)
the solution of system (\ref{12}) exists on any interval $t\in[0, \ T]$
such that the smooth solution~$u_0$ of the Hopf equations exists on
this interval.

    System (\ref{12}) can be solved in the following way: first, one finds
the smooth solution of the Hopf equation, next, one finds the function
$e(x,t)$ from the last equations (which is uniquely solvable in view
of the inequality $2u_0(\phi,t)<\phi_t$), then one finds the (positive)
function~$g(t)$ from the next to the last equation, and finally, one
finds the function $\phi(t)$.

    Note that system (\ref{12}) contains no obstacles to setting
$e(x,t)= 0$. If so, $g(t)={\mbox{const}}$ in the case of an
arbitrary (nonconstant) background function $u_0(x,t)$. But this
conclusion is contrary to well known properties of soliton
solutions of the KdV equation (see, e.g.,~\cite{3}).

    Moreover, under our notation, the weak asymptotics of the
asymptotic one-soliton solution to the KdV equation, constructed
by V.~P.~Maslov and G.~A.~Omelyanov~\cite{3}, has the form
\begin{equation}
\label{13}
u^*_{1,\varepsilon }(x,t)= u_{01}(x,t)
+g_1(t){\varepsilon }\delta(x-\phi_1(t))
+ e_1(x,t){\varepsilon }[1-\theta(x-\phi_1(t))], \quad \varepsilon  \to+0.
\end{equation}
In other words, in the case (\ref{8}) the "shock wave" with a
small amplitude ${\varepsilon }e(x,t)\theta(x-\phi_1(t))$
propagates {\it in front of the soliton\/}
${\varepsilon }\delta(x-\phi_1(t))$, but in the asymptotic one-soliton
solution constructed in~\cite{3} the small shock wave
${\varepsilon }e_1(x,t)[1-\theta(x-\phi_1(t))]$ arises
{\it behind the soliton \/}.

    If we apply Definition (\ref{10}) or (\ref{11}) to the
asymptotic solution obtained in~\cite{3}, whose weak asymptotics
yields (\ref{13}), we obtain the following system of equations
\begin{equation}
\label{14}
\begin{array}{rcl}
\displaystyle
u_{01t}+(u^2_{01})_x &=&0,\\
\displaystyle
\phi_{1t} - 2u_{01}(\phi_1(t),t) - \frac{2}{3}g_1(t) &=&0, \\
\displaystyle
                                                      &&\\
\displaystyle
e_1(\phi(t),t)+\frac{3\sqrt{6}}{2}g_{1t}(t)/g^{3/2}_1(t) &=&0,\\
\displaystyle
\big(e_{1t}(x,t) + 2(u_{01}(x,t)e_1(x,t))_x \big)\Big|_{x < \phi_1(t)}&=&0.
\end{array}
\end{equation}

    The solution of the last system for $g_{1t}(t)\ne 0$ is not
uniquely determined by the initial conditions $e_1(x,0)$ for
$x \le \phi_1(0)$, since the velocity along the characteristic
($\dot x= 2u_{01}(x(t),t)$) is less (for $g_{1}(t)>0$) than that
of the soliton $\phi_{1t}=2u_{01}(\phi_1(t),t)+\frac{2}{3}g_1(t)$.

    Thus, the assumption that the structure of the solution to the KdV
equation is specified by (\ref{13}) due to Definitions (\ref{10}),
(\ref{11}) leads to an ill-posed Cauchy problem (with a nonunique
solution) for the functions $u_{01}(x,t)$, \ $g_1(t)$, \ $e_1(x,t)$,
\ $\phi_1(t)$.

    On the other hand, the system of equations obtained in~\cite{3} for
these functions has the form
\begin{equation}
\label{15}
\begin{array}{rcl}
\displaystyle
u_{01t}+(u^2_{01})_x &= &0,\\
\displaystyle
\phi_{1t} - 2u_{01}(\phi_1(t),t) - \frac{2}{3}g_1(t) &=&0, \\
\displaystyle
e_1(\phi(t),t)+\frac{3\sqrt{6}}{2}g_{1t}(t)/g_1^{3/2}(t) &=&0,\\
\displaystyle
\big(e_{1t}(x,t) + 2(u_{01}(x,t)e_1(x,t))_x \big)\Big|_{x<\phi_1(t)}&=&0,\\
\displaystyle
g_{1}(t) + 2u_{01}(\phi_1(t),t) &= &{\mbox{const}}, \\
\end{array}
\end{equation}

    It is evident that this system differs from system (\ref{14}) obtained
from (\ref{10}), (\ref{11}), (\ref{13}) by the additional equation
$g_{1}(t)+2u_{01}(\phi_1(t),t)=g_{1}(0)+2u_{01}(\phi_1(0),0)$.
The presence of this equation implies that system (\ref{15}) splits into
the two systems
\begin{equation}
\label{16}
\begin{array}{rcl}
\displaystyle
u_{01t}+(u^2_{01})_x &= &0,\\
\displaystyle
\phi_{1t} - 2u_{01}(\phi_1(t),t) - \frac{2}{3}g_1(t) &= &0, \\
\displaystyle
g_{1}(t) + 2u_{01}(\phi_1(t),t) &= &{\mbox{const}}, \\
\end{array}
\end{equation}
and
\begin{equation}
\label{17}
\displaystyle
\big(e_{1t}(x,t) + 2(u_{01}(x,t)e_1(x,t))_x \big)\Big|_{x<\phi_1(t)}=0,
\end{equation}

\begin{equation}
\label{18}
\displaystyle
e_1(\phi(t),t)+\frac{3\sqrt{6}}{2}g_{1t}(t)/g_1^{3/2}(t) = 0,
\end{equation}
and equality (\ref{18}) is the boundary condition for equation (\ref{17}),
which turns the Cauchy problem for equation (\ref{17}) into the well-posed
one (the Cauchy condition, in view of (\ref{13}), has the form
$e_1(x,0)= e_1^0(x)[1-\theta(x-\phi_1(0))]$).

    Moreover, if equation (\ref{17}) is considered formally in the
domain $x>\phi_1(t)$, which corresponds to the solution structure given by
formula (\ref{8}), then the "redundant" condition
$$
e(\phi(t),t)-\frac{3\sqrt{6}}{2}g_{t}(t)/g^{3/2}(t) = 0,
$$
analogous to (\ref{18}), overdetermines the problem.

     Thus, the weak asymptotics corresponding to the asymptotic solution
of the Cauchy problem for the KdV equation constructed in~\cite{3} cannot
be derived from the solution to the KdV equation with the help of
Definitions (\ref{10}) or (\ref{11}), and vice versa.

    Why is it so? The essence of the matter lies in the definition of
weak (generalized) solution to nonlinear equation. It turns out that
the definition of the weak (generalized) solution to nonlinear equation
depends on the structure of the kernel of the operator adjoint to
the linearized operator of the initial differential equation which
arises when constructing the smooth asymptotics. This construction
of the definition of weak solutions was previously discussed
in~\cite{17},~\cite{18}.

    In the present paper we do not come into details of construction of
the definition of the weak solution to our problem. We just point out
that, in terms of this construction, the KdV equation is analogous to
the phase field system discussed in~\cite{18}.

    The difference is that for the KdV equation the kernel of
the adjoint operator mentioned above is {\it two-dimensional\/},
which results in the following definition analogous to that of
the weak solution in the form of the integral identity from~\cite{18}.

\begin{de}
\label{de1} \rm
    The asymptotic distribution of the form (\ref{13})
$$
u^*_{\varepsilon }(x,t)= u_0(x,t)+g(t){\varepsilon }\delta(x-\phi(t))
+ e(x,t){\varepsilon }\theta(-x+\phi(t))
$$
is a generalized asymptotic (soliton-type) solution to the KdV equation
(\ref{1}) for $t \in [0, T]$ with the initial condition
$u^{0*}_{\varepsilon }(x)$, if for any constants $c_1$, \ $c_2$ the
following equality holds
\begin{equation}
\label{19}
\begin{array}{rcl}
\displaystyle
\big(c_1+ c_2u^*(x,t,\varepsilon )\big)L_{KdV}[u^*(x,t,\varepsilon )]&=&
O_{{\cal D}'}(\varepsilon ^2), \\
\displaystyle
u^*_{\varepsilon }(x,0) &= & u^{0*}_{\varepsilon }(x)
+O_{{\cal D}'}(\varepsilon ^2), \\
\end{array}
\end{equation}
where $u^*(x,t,\varepsilon )$ is a smooth approximation of the
asymptotic distribution $u^*_{\varepsilon }(x,t)$, and the first
estimate is uniform with respect to $t \in [0, T]$.
\end{de}

    It is clear that (\ref{19}) is equivalent to the following relations
$$
L_{KdV}[u^*(x,t,\varepsilon )] = O_{{\cal D}'}(\varepsilon ^2),
\qquad
u^*(x,t,\varepsilon )L_{KdV}[u^*(x,t,\varepsilon )]
= O_{{\cal D}'}(\varepsilon ^2),
$$
and the first relation coincides with Definition (\ref{10}).

    One can easily see that (\ref{19}) can be rewritten as an integral
identity but of an unusual form.

    By analogy to what was previously said, the solution depends on
the choice of the approximation, and to obtain the results known in
the theory of the KdV equation one should choose, as an approximation
of the asymptotic distribution $g(t){\varepsilon }\delta(x-\phi(t))$,
the function from the formula for asymptotic solution to the KdV
equation~\cite{3}:
$$
\displaystyle
g(t){\varepsilon }\delta(x-\phi(t),\varepsilon )
= g(t)\omega\big(\alpha(t)\frac{x-\phi(t)}{\varepsilon }\big),
$$
where $\alpha(t)= \sqrt{\frac{g(t)}{6}}$, \ $\omega(z)= \ch^{-2}(z)$.

    It is well known that the function $\omega(\alpha(t)\tau)$ is a
solution of the boundary value problem for the differential equation
\begin{equation}
\label{20}
\displaystyle
-\phi_t(t)\frac{d\omega}{dz}+2(u_0(\phi(t),t)+g(t)\omega)\frac{d\omega}{dz}
+\alpha^2(t)\frac{d^3\omega}{dz^3}= 0,
\end{equation}
where $\omega(z) \to 0$ as $|z|\to \infty$.

    It is easy to see that this boundary value problem has the solution
given above provided that
$$
\phi_{t}=2u_0(\phi(t),t) + \frac{2}{3}g(t).
$$

    In fact, the equation for the approximation~$\omega$ can be derived
by applying an analog of the average procedure, but in the present paper
we shall simply postulate the choice of approximation (\ref{20}).

    In Section~\ref{s3} it is shown that for approximation of a
soliton of the form (\ref{20}), Definition~\ref{de1} immediately
leads to the system of equations (\ref{15}) previously obtained
in~\cite{3} when constructing the smooth asymptotics of
one-soliton solution to the KdV equation.

    It should be emphasized that Definition~\ref{de1} for the Hopf equation
is {\it not equivalent\/} to that for the KdV equation since the term
${\varepsilon }^2u_{xxx}$, for $c_2\ne 0$, makes a contribution of the
order of $O_{{\cal D}'}(\varepsilon )$ to the left-hand side of the equality.

    Thus, it can be said that {\it asymptotic generalized solutions exist
for both the Hopf and KdV equations \/}. When $u_0(x,t)= {\mbox const}$,
the systems describing this solution {\it coincide \/} but, in the general
case ($u_0(x,t)\ne {\mbox const}$), they prove to be absolutely
{\it distinct \/} in the sense that the limit problem for any one of
them is ill-posed for the other.

    Moreover, the formal $\delta$-soliton generalized asymptotic solution
of the Hopf equation is not related (is not an asymptotics) to
any exact solution of the Hopf equation. In order to explain this fact,
we construct the exact solution of this equation satisfying the Cauchy
condition
$$
W(x,t,\varepsilon )\big|_{t=0}=\varepsilon \delta(x).
$$
For $t>0$ this (discontinuous) solution has the form
$$
W(x,t,\varepsilon )=\left\{
\begin{array}{ll}
0,&\quad x\leq 0,\\
x/2t, &\quad 0\leq x\leq 2\sqrt{\varepsilon t},\\
0,&\quad x>2\sqrt{\varepsilon t}.
\end{array}
\right.
$$
The discontinuity points
$\phi(t)= 2\sqrt{\varepsilon t}$ can be calculated from the
Rankine--Hugoniot conditions. Let $\varphi(x)\in{\cal D}(\RA'_x)$
be an arbitrary test function. Let us calculate the limit
$$
\lim_{t\to+0}\langle W(x,t,\varepsilon ),\varphi(x)\rangle
= \lim_{t\to+0}\frac1{2t}\int^{2\sqrt{\varepsilon t}}_0x\varphi(x)\,dx.
$$
By using the L'Hospital rule, we obtain
$$
\lim_{t\to+0}\langle W(x,t,\varepsilon),\varphi(x)\rangle
= \varepsilon \varphi(0).
$$
It is clear that for $t=1$ the support of~$W(x,t,\varepsilon)$
is the interval $[0,2\sqrt{\varepsilon}]$, while for $t=t_0$ the
singularity support of the generalized $\delta$-soliton constructed,
in view of Definition~(\ref{10}), with an {\it arbitrary\/} approximation
$\omega(x/\varepsilon)$ of the initial value $\varepsilon\delta$
can be found at the point
$$
\phi(t_0)= t_0\int \omega^2(z)\,dz,
$$
which lies outside the support of the exact solution $W(x,t,\varepsilon)$
(the method used for deriving this relation will be discussed below).

    The results, similar to mentioned above can be obtained for the KdV
type equation $u_t+(f(u))_x+\varepsilon^2u_{xxx}= 0$ with arbitrary smooth
nonlinearity, which admits soliton solutions. To verify this assertion,
by analogy to~(\ref{20}), we consider the boundary value for the ordinary
differential equation
$$
-\phi_t\frac{d\omega}{d\tau} + \frac{d}{d\tau}f(u_0(\phi,t)+\omega)
+\frac{d^3}{d\tau^3}\omega= 0,
$$
where $u_0$, \ $\phi$ are treated as parameters and the solution
$\omega=\omega(\tau,t)$ is chosen as an approximation of the term
$\varepsilon g(t)\delta\big(x-\phi(t)\big)$ in the singular
ansatz. Integrating this equation, multiplying the obtained equation by
$\omega_{\tau}$, and integrating again, we obtain the energy
conservation law
$$
\frac{1}{2}\big(\omega_{\tau}\big)^2 + \Phi(\omega)= E,
$$
where
the function $\Phi(\omega,t)=\int^\omega_0 f(z+u_0)\,dz
-\frac{1}{2}\phi_t\omega^2$ is considered as a potential and $E$
is an arbitrary constant having the meaning of energy.

    Integrating the last equation, we find
\begin{equation}
\label{23}
\tau -\tau_0=\pm\int_{\omega_1}^{\omega}\frac{d\xi}{\sqrt{E-\Phi(\xi,t)}},
\end{equation}
where $\omega_1$, \ $\omega_2$ are the last two roots of the equation
$\Phi(\xi)= E$, the radicand is assumed to be positive when
$\omega_1 < \omega(\tau) < \omega_2$, \ $\tau_0$ is an arbitrary constant.

    Using the standart methods, one can show that, for a certain
choice of the function $f(u)$ and the constant $E$,
equation (\ref{23}) has soliton solutions (see~\cite[\S 1]{20}).

    For the existence of localized solution it is necessary that
one of the roots $\omega_1$, \ $\omega_2$ of the equation
$\Phi(\xi)= E$ is multiple (that is, the integral in (\ref{23}) is
divergent).

    Thus, in the case of equations with general nonlinearity, we shall
substitute into them the smooth ansatzs
\begin{equation}
\label{24}
u^*(x,t,\varepsilon )=u_0(x,t)
+ \omega\big(\frac{x-\phi(t)}{\varepsilon },t\big)
+ e(x,t){\varepsilon }\omega_0(\frac{x-\phi(t)}{\varepsilon }),
\quad \varepsilon  >0,
\end{equation}
where the function $\omega(\tau,t)$ decreases
sufficienly rapidly with respect to $\tau$ and
$\omega_0({x}/{\varepsilon })$ is an approximation of the Heaviside
function $\theta(x)$.

    Accordingly, in this case the weak asymptotics (\ref{24}), that is,
for equations of Hopf or KdV type, the singular ansatz has the form
\begin{equation}
\label{25}
u^*_{\varepsilon }(x,t)= u_0(x,t)
+\Omega_1(t){\varepsilon }\delta\big(x-\phi(t)\big)
+ e(x,t){\varepsilon }\theta\big(\pm (x-\phi(t))\big),
\quad \varepsilon  \to +0,
\end{equation}
$$
\Omega_1(t)= \int\omega(\tau,t)\,d\tau.
$$

    Note that, generally speaking, we need not use any fixed approximations
for the distributions contained in~(\ref{25}). In this case, we obtain
equations that describe the dynamics and, in general, depend on the
approximation of singular generators of the ansatz.

    The dependence of the asymptotic discontinuous solution to a
nonlinear equation on approximations of singular generators of
the solution is substantial for our approach. It is due to the
fact that asymptotics of the product of approximations is
independent of the choice of these approximations only in
specific cases, for example, when the factors are "weakly"
singular, as when calculating the asymptotics of the product of
the aproximation of a distribution by the approximation of a
smooth function. Naturally, there is no such dependence if we
use this method to find discontinuous solutions of linear
equations.

    By what has been said, when constructing the set of
asymptotic distributions in which singular ansatzs of the form
(\ref{8}), (\ref{13}) are embedded all possible approximations
will be considered for each distribution. The same idea is basic
to the construction of the Colombeau algebra of new generalized
functions~\cite{6},~\cite{7}. In the Colombeau theoty there are
also "many" $\delta$-functions associated with "one" Schwartz's
$\delta$-function.

    4. Let us summarize the main results. One of the most important
results is Definition~\ref{de1} for generalized asymptotic solutions
of KdV type equations. In order to explain how to use this defintion,
in Section~\ref{s2} of the present paper we construct an associative
algebra of smooth functions ${\cal H}^*$ in which the smooth ansatzs
(\ref{10}), (\ref{14}) or (\ref{24}) are embedded which are
approximations of the singular ansatzs (\ref{12}), (\ref{15}) or
(\ref{25}). The singular ansatzs on which generalized infinitely
narrow soliton-type solutions are constructed belong to the set
of asymptotic distributions ${\cal E}^*$ and are derived as weak
asymptotics of elements from ${\cal H}^*$. In this section we
also give a definition of generalized solutions to nonlinear
equations, that is, rules of substituting singular ansatzs into
these equations.

    In Section~\ref{s3} we derive the above-mentioned systems of
equations (1.12) and (1.15) by using different definitions of
the generalized solution.

    In Section~\ref{s4} we write out the system of equations
determining the dynamics of the infinitely narrow
$\delta$-soliton to equations of Hopf and KdV types with general
nonlinear terms.

\section{Asymptotic distributions and generalized solutions to nonlinear
equations}%2
\label{s2}

    1. When solving problems listed in Introduction it is necessary to
construct a singular solution with singular generators of the special
type (\ref{8}) or (\ref{13}) to the nonlinear equation $L[u]=0$. According
to our method of weak asymptotics, we substitute into the equation
$L[u]=0$ smooth ansatzs of the form (\ref{9}) resulting from the
replacement of singular generators by their approximations in singular
ansatzs of the form (\ref{8}).

    It is well known~\cite[ch.I, \S 4.6]{22} that to each distribution
(generalized function) $f(x)\in {\cal D}'$ can be assigned its approximation
\begin{equation}
\label{30}
f(x,\varepsilon )= f(x)*K(x,\varepsilon )=
\langle f(t), K(x-t,\varepsilon )\rangle, \quad \varepsilon  >0,
\end{equation}
where $*$ is a convolution, the kernel
$K(x,\varepsilon )= \frac{1}{\varepsilon }
\omega\big(\frac{x}{\varepsilon }\big)$ is a $\delta$-type function such
that $\omega(z)\in C^{\infty}(\RA)$, \ $\omega(z)$ has a compact support or
decreases sufficiently rapidly as $|z|\to \infty$, for example,
$|\omega(z)|\le C(1+|z|)^{-3}$ and $\int\omega(z)\,dz=1$.

    For all test functions $\varphi(x) \in {\cal D}$ we have:
$$
\lim\limits_{\varepsilon  \to +0}\langle f(x,\varepsilon ),\varphi(x)\rangle
=\langle f(x),\varphi(x) \rangle.
$$

    We cite the approximations of the distributions $\delta(x)$,
\ $\theta(x)$,
which are singular generators of the soliton-type ansatzs (\ref{8}),
(\ref{13}).

    For the approximation of $\delta$-function we have from (\ref{30}):
\begin{equation}
\label{31}
\delta(x,\varepsilon )= \frac{1}{\varepsilon }\omega(\frac{x}{\varepsilon }).
\end{equation}

    For the approximation of the Heaviside function $\theta(x)$ we have
$$
\theta(x,\varepsilon )= \theta(x)*\frac{1}{\varepsilon }
\omega\big(\frac{x}{\varepsilon }\big)
=\int_{0}^{\infty}\omega\big(\frac{x}{\varepsilon }-t\big)\,dt.
$$
Hence we find
\begin{equation}
\label{32}
\theta(x,\varepsilon )= \omega_0(\frac{x}{\varepsilon }),
\end{equation}
where $\omega_0(z)= \int_{-\infty}^{z}\omega(\eta)\,d\eta$, \
$\lim_{z \to +\infty}\omega_0(z)= 1$, \ $\lim_{z \to-\infty}\omega_0(z)=0$,
\ $\omega_0(z)\in C^{\infty}(\RA)$.

    If the Cauchy kernel $K(x,\varepsilon )=
\frac{1}{\pi}\frac{\varepsilon }{x^2 + {\varepsilon }^2}$ is used in
(\ref{30}), we obtain harmonic approximations
$f(x,\varepsilon )$ for distributions $f(x)$. In this case
(see~\cite{22},~\cite{23}):
$$
\delta^{(m-1)}(x,\varepsilon )
= \frac{(-1)^m (m-1)!}{2 \pi i}(z^{-m} - {\overline{z}}^{-m}),
\qquad \theta(x,\varepsilon )=\Big(1+\frac{2}{\pi}
\arctg\big(\frac{x}{\varepsilon }\big)\Big),
$$
where $z= x+i\varepsilon $, \ ${\overline{z}}= x-i\varepsilon $, \
$m=0,1,2,\dots$.

    Denote by ${\cal H}^*$ the associative and commutative differential
algebra generated by finite sums of finite products of functions
$f(x,\varepsilon )$ from the space ${\cal H}_0$ of smooth ansatzs
$u^*(x,t,\varepsilon )$ of the form (\ref{9}), where approximations
of the $\delta$-function and the Heaviside function are defined in
(\ref{31}) and (\ref{32}), respectively.

    Denote by ${\cal E}^*$ the set of weak asymptotics
$f^*_{\varepsilon }(x,t)$ derived from elements $f^*(x,t,\varepsilon )$
of the algebra ${\cal H}^*$, as $\varepsilon  \to +0$, determined up to
$O_{{\cal D}'}(\varepsilon ^2)$, and, according to~\cite{11},~\cite{14},
call it the set of {\it asymptotic (soliton-type) distributions\/}.

    By analogy with~\cite{14}, one can introduce a structure of the
asymptotic algebra on ${\cal E}^*$ which is not, however, required
in the present paper.

    Thus, for example, the one-soliton solution to the KdV equation
(\ref{2}) belongs to the algebra ${\cal H}^*$ and its weak asymptotics
(\ref{3}) and singular ansatzs of the form (\ref{18}) are asymptotic
distributions from ${\cal E}^*$.

    It can be shown (see~\cite{14}) that each element
$f^{*}(x,t,\varepsilon )$ from the algebra ${\cal H}^*$ has a weak
asymptotics of the form
\begin{equation}
\label{33}
f^{*}(x,t,\varepsilon )=f_0(x,t)
+f_1(x,t){\varepsilon }\delta\big((x-\phi(t))^{-1}\big)
+f_2(x,t){\varepsilon }\theta\big((x-\phi(t))^{-1}\big)
+ O_{{\cal D}'}(\varepsilon ^2), \quad \varepsilon  \to +0,
\end{equation}
where $f_j(x,t)$ are smooth functions, $j=0,1,2$.

    Finding weak asymptotics of the elements from the algebra
${\cal H}^*$ can be interpreted, in a sense,
as multiplications of asymptotic distributions
(distributions)
(see~\cite{10},~\cite{11} and ~\cite{12}--~\cite{14}).

    Let us give some examples. The following equalities can be
readily verified:
\begin{equation}
\label{34}
\begin{array}{rcl}
\displaystyle
[\varepsilon \delta(x,\varepsilon )]^n &=& \Omega_n\varepsilon \delta(x)
+ O_{{\cal D}'}(\varepsilon ^2),\\
\displaystyle
\varepsilon \theta(x,\varepsilon ) &=& \varepsilon \theta(x)
+ O_{{\cal D}'}(\varepsilon ^2),\\
\displaystyle
[\varepsilon \theta(x,\varepsilon )]^m &=& O_{{\cal D}'}(\varepsilon ^m),\\
\displaystyle
[\varepsilon \delta(x,\varepsilon )]^n
[\varepsilon \theta(x,\varepsilon )]^{m-1}
&=& O_{{\cal D}'}({\varepsilon }^{m}). \\
\end{array}
\end{equation}
where $n=1,2,\dots$, \ $m=2,\dots$, \ $\omega(z)$ is a $\delta$-type
function from (\ref{30}), $\Omega_n=\int_{-\infty}^{\infty}\omega^n(z)\,dz$.

    In particular, it follows herefrom that the asymptotic distributions
$\varepsilon \delta(x)$ and $\varepsilon \theta(x)$ constitute
an algebra, which is asymptotic $\mod O_{{\cal D}'}(\varepsilon ^2)$.

   2. Let $F(u)$ be a smooth function of at most exponential growth. We
define a function of asymptotic distribution
$F\big(f^*_{\varepsilon }(x,t)\big)\in {\cal E}^*$ as the weak asymptotics
of the function $F\big(f^*(x,t,\varepsilon )\big)$, as
$\varepsilon  \to +0$, where $f^{*}(x,t,\varepsilon )\in {\cal H}^*$
is the approximation of the asymptotic distribution
$f^*_{\varepsilon }(x,t)\in {\cal E}^*$.

   Then, for example, $F\big(\varepsilon \delta(x)\big)$ is defined
as the weak asymptotics of the approximating function
$F\Big(\omega\big(\frac{x}{\varepsilon }\big)\Big)$ as $\varepsilon  \to+0$.
Taking into account the estimate for the function $\omega$ and using the
Lagrange theorem
$$
F(\omega(z))-F(0)= F'(\Theta\omega(z))\omega(z), \quad 0<\Theta <1,
$$
we have
$$
|F(\omega(z))-F(0)| \le K(1+|z|)^{-3}.
$$

    Thus, for the function $F(\omega(z))-F(0)$ there exists an
estimate analogous to that for the function $\omega(z)$
approximating the asymptotic distribution $\varepsilon
\delta(x)$. Therefore, applying
$F\Big(\omega\big(\frac{x}{\varepsilon }\big)\Big)$ to a test
function and performing the change of variables $x= \varepsilon \eta$,
we obtain
$$
J(\varepsilon )= \langle F(0),\varphi(x)
\rangle +
\Big\langle F\Big(\omega\big(\frac{x}{\varepsilon }\big)\Big)-F(0),
\varphi(x) \Big\rangle
$$
$$
= \langle F(0),\varphi(x)\rangle + \varepsilon \int_{-\infty}^{\infty}
\Big[F\Big(\omega\big(\eta\big)\Big)-F(0)\Big]
\varphi(\varepsilon \eta)\,d\eta
$$
$$
= \langle F(0),\varphi(x) \rangle +
\varepsilon \varphi(0)\int_{-\infty}^{\infty}
\Big[F\Big(\omega\big(\eta\big)\Big)-F(0)\Big]\,d\eta
+ O(\varepsilon^2 ), \quad \varepsilon  \to +0.
$$

    Thus, in the weak sense, we have
\begin{equation}
\label{35}
F\big(\varepsilon \delta(x)\big)= F(0)+ \varepsilon \Lambda\delta(x)
+ O_{{\cal D}'}(\varepsilon^2 ), \quad \varepsilon  \to +0,
\end{equation}
where the constant
$\Lambda= \int\Big[F\big(\omega(\eta)\big)-F(0)\Big]\,d\eta$.

\section{Infinitely narrow solitons to the Hopf and KdV equations}%3
\label{s3}

    1. {\it The Hopf equation: $u_t + (u^2)_x=0$.} To find an
asymptotic $O_{{\cal D}'}(\varepsilon ^2)$ infinitely narrow
$\delta$-soliton-type solution to the Hopf equation (\ref{4}) of the form
\begin{equation}
\label{45}
u^*_{\varepsilon }(x,t)=u_0(x,t) + g(t)\varepsilon \delta(x-\phi(t)),
\quad \varepsilon  \to +0,
\end{equation}
where $u_0(x,t)\in C^{\infty}({\RA}^2)$, $g(t), \phi(t)\in C^{\infty}(\RA)$
are the desired functions, one needs to substitute into the
equation its approximation from ${\cal H}^*$ of the form:
\begin{equation}
\label{46}
u^*(x,t,\varepsilon )=u_0(x,t)+g(t)\varepsilon \delta(x-\phi(t),\varepsilon ),
\quad \varepsilon >0,
\end{equation}

where, according to (\ref{31}), the function $\varepsilon
\delta(x,\varepsilon )= \omega(\frac{x}{\varepsilon })$ is used
as an approximation of the asymptotic distribution $\varepsilon \delta(x)$.

    Using the first formula (\ref{34}) for the squared asymptotic
distribution $\varepsilon \delta(x)$, we obtain
$$
[u^*_{\varepsilon }(x,t)]^2= u_0^2 + 2u_0g\varepsilon \delta(x-\phi(t))
+ \varepsilon  g^2\Omega\delta(x-\phi(t))
+ O_{{\cal D}'}(\varepsilon ^2), \quad \varepsilon  \to +0.
$$
    Then, substituting $u^*_{\varepsilon }(x,t)$ into the Hopf equation
we have
$$
 u^*_{\varepsilon  t} + [(u^*_{\varepsilon })^2]_x = u_{0t}+(u^2_0)_x
+\varepsilon \delta(x-\phi)\Big(g_t+2g u_{0 x}\Big)
$$
\begin{equation}
\label{47}
+ \varepsilon \delta'(x-\phi(t))\Big(-g\phi_t+ \Omega g^2+2u_0g\Big)
+ O_{{\cal D}'}(\varepsilon ^2), \quad \varepsilon  \to +0,
\end{equation}
where $\Omega= \int\omega^2(\eta)\,d\eta$.

    Using the well-known equality
\begin{equation}
\label{48}
a(x)\delta'(x) = a(0)\delta'(x) - a'(0)\delta(x),
\end{equation}
and equating the coefficients of $\varepsilon ^0$, \ $\varepsilon \delta$
and $\varepsilon \delta'$ with zero, we obtain the necessary and
sufficient conditions for the right-hand side of equation (\ref{47}) to
be of the order of $O_{{\cal D}'}(\varepsilon ^2)$:
\begin{equation}
\label{49}
\begin{array}{rcl}
\displaystyle
u_{0t}+(u^2_0)_x &=&0,\\
\displaystyle
\phi_t - \Omega g(t) - 2u_0(x,t)\Big|_{x= \phi(t)} &=&0, \\
\displaystyle
g_t(t) &=& 0. \\
\end{array}
\end{equation}

    In the case of constant background $u_0= \mbox{const}$ we have
$\phi(t)= vt + \phi_0$, where $v= \Omega g(0) + 2u_0$ and $\phi_0$
is a constant which has the meaning of the coordinate of the initial
position of the soliton.

    It is clear that the condition $g_t= 0$, when $u_0\ne\mbox{const}$,
contradicts the physical intuition and the well-known results about
soliton behaviour. However, formally, such a structure can also exist,
since system~(\ref{53}) is well-defined, but we can try to repair the
situation  by adding a new term which, after differentiation,
has the form ${\varepsilon }\delta(x-\varphi)$. Therefore, we shall
seek an asymptotic solution to the Hopf equation of the form (\ref{8})
by substituting into the equation the smooth ansatz of the type (\ref{9}):
\begin{equation}
\label{50}
u^*(x,t,\varepsilon )= u_0(x,t)
+ g(t)\omega\big(\alpha(t)\frac{x-\phi(t)}{\varepsilon }\big)
+ e(x,t){\varepsilon }\omega_0(\frac{x-\phi(t)}{\varepsilon }),
\quad \varepsilon  >0,
\end{equation}
where $\alpha(t)\in C^{\infty}$, \ $\Omega_1= \int\omega(\eta)\,d\eta$, \
$\omega(z)$ is a $\delta$-type function which has a compact support or
decreases sufficiently rapidly as $|z|\to \infty$, \
$\omega_0(\frac{x}{\varepsilon })$ is an approximation of the Heaviside
function.

   The weak asymptotics (\ref{50}) has the form
\begin{equation}
\label{51}
u^*_{\varepsilon }(x,t)= u_0(x,t)
+g(t)\Omega_1{\varepsilon }\delta\big(\alpha(t)(x-\phi(t))\big)
+ e(x,t){\varepsilon }\theta(x-\phi(t)), \quad \varepsilon  \to +0.
\end{equation}
This representation coincide with one that can be obtained when
considering the smooth asymptotics of the asymptotic solution to the
KdV equation with a small dispersion $\sim {\varepsilon }^2$ ~\cite{1},
~\cite{3}, in the weak sense as $\varepsilon  \to +0$.

    In this case, according to Section~\ref{s2},
$$
[u^*_{\varepsilon }(x,t)]^2=u_0^2(x,t)
+ \frac{1}{\alpha(t)}\Big(2u_0(x,t)g(t)\Omega_1 + g^2(t)\Omega_2\Big)
{\varepsilon }\delta(x-\phi(t))
$$
\begin{equation}
\label{52}
+ 2u_0(x,t)e(x,t){\varepsilon }\theta(x-\phi(t))
+ O_{{\cal D}'}(\varepsilon ^2), \quad \varepsilon  \to +0,
\end{equation}
where $\Omega_1= \int\omega(\eta)\,d\eta$, \
$\Omega_2= \int\omega^2(\eta)\,d\eta$.

   Substituting (\ref{50}) into the Hopf equation and using (\ref{51}),
(\ref{52}), up to the terms of the order of $O_{{\cal D}'}({\varepsilon }^2)$,
we obtain a relation analogous to (\ref{47}). Then, setting the
coefficients of ${\varepsilon }^0$, \
${\varepsilon }\theta(x-\phi(t))$, \ ${\varepsilon }\delta(x-\phi(t))$
and ${\varepsilon }\delta'(x-\phi(t))$ equal to zero, we arrive at the
following result.

\begin{th}
\label{th2}
    Suppose that for $t \in [0, \ T]$ there exists a smooth solution
$u_0(x,t)$ to the Hopf equation with the smooth initial condition
$u_0(x,t)\Big|_{t= 0}= u_0^0(x)$.
    Then, on the closed interval $[0, \ T]$, the Hopf equation
has a solution in the form of an infinitely narrow
$\delta$-soliton
{\rm (\ref{51})}  if and only if the unknown smooth
functions $u_0(x,t)$, $g(t)$, \ $e(x,t)$, \ $\alpha(t)$, \ $\phi(t)$
satisfy the following system of equations:
\begin{equation}
\label{53}
\begin{array}{rcl}
\displaystyle
u_{0t}+(u^2_0)_x &= &0,\\
\displaystyle
\phi_{t} - 2u_0(\phi(t),t) - \frac{\Omega_2}{\Omega_1}g(t) &=&0, \\
\displaystyle
e(\phi(t),t)-\frac{\Omega_1^2}{\Omega_2}\frac{(g(t)/\alpha(t))_t}{g(t)}&=&0,\\
\displaystyle
\big(e_{t}(x,t) + 2(u_0(x,t)e(x,t))_x \big)\Big|_{x > \phi(t)} &=&0,
\end{array}
\end{equation}
where $\Omega_1=\int\omega(\eta)\,d\eta$, \
$\Omega_2=\int\omega^2(\eta)\,d\eta$.
\end{th}

    {\bf Remark.} System (\ref{12}) studied in the Introduction can
be obtained from system (\ref{53}) if we take the solution of equation
(\ref{20}) as the approximation $\omega(\alpha(t)(x-\varphi(t))/\varepsilon)$.
In this case we have
$$
\omega= \cosh^{-2}(\tau),\qquad \alpha= \sqrt{g(t)/6}
$$
and system (\ref{53}) turns into systems (\ref{12}), where the number of
unknown functions is equal to the number of equations.

    2. {\it The KdV equation: $u_t + (u^2)_x + \varepsilon ^2u_{xxx}=0$.}
This equation has the exact one-soliton solution (\ref{2})
$$
u(x,t,\varepsilon )
=g\ch^{-2}\Big(\sqrt{\frac{g}{6}}\big(x-\frac{2}{3}gt\big)/\varepsilon \Big),
$$
where $g=\mbox{const}$ is the amplitude of the soliton~\cite{25}. The weak
asymptotics of this solution has the form (\ref{3})
$$
u(x,t)=g{\varepsilon }\delta(x-\frac{2}{3}gt), \quad \varepsilon  \to +0.
$$

   Let us study the dynamics of propagation of an infinitely narrow deformed
soliton solution of the KdV equation. To this end, we consider a smooth
ansatz of the form
\begin{equation}
\label{54}
u^*(x,t,\varepsilon )= u_0(x,t)
+g(t)\omega(\alpha(t)\frac{x-\phi(t)}{\varepsilon })
+ e(x,t){\varepsilon }\omega_0^{-}(\frac{x-\phi(t)}{\varepsilon }),
\quad \varepsilon  >0,
\end{equation}
where $\alpha(t)= \sqrt{\frac{g(t)}{6}}$, \ $\omega(\eta)= \ch^{-2}(\eta)$, \
$\omega_0^{-}(\frac{x}{\varepsilon })= 1-\omega_0(\frac{x}{\varepsilon })$,
and, according to (\ref{32}),
$$
\omega_0(\frac{x}{\varepsilon })
= \int_{-\infty}^{x/\varepsilon }\omega_1(\eta)\,d\eta.
$$
Here $\omega_0(\frac{x}{\varepsilon })$ and
$\omega_0^{-}(\frac{x}{\varepsilon })$ are approximations of the Heaviside
functions $\theta(x)$ and $\theta(-x)$, respectively,
$\int\omega_1(\eta)\,d\eta=1$.

   The weak asymptotics of the right-hand side of (\ref{54}) has the form
(\ref{51}).

    According to Definition~\ref{de1}, the infinitely narrow soliton-type
solution of the KdV equation is defined as the solution of the following
system of the two equations
\begin{equation}
\label{55}
\begin{array}{rcl}
\displaystyle
L_{KdV}[u]= u_t + (u^2)_x + {\varepsilon }^2u_{xxx} &=&
O_{{\cal D}'}(\varepsilon^2), \\
\displaystyle
uL_{KdV}[u]=(u^2)_t + \frac{4}{3}(u^3)_x + {\varepsilon }^22uu_{xxx}&=&
O_{{\cal D}'}(\varepsilon^2).
\end{array}
\end{equation}
Let us write the dispersion term of the second equation in the form
${\cal G}(u,\varepsilon )= {\varepsilon }^22uu_{xxx}
={\varepsilon }^2[(u^2)_{xx}-3(u_x)^2]_x$.

     Up to the terms whose asymptotics are of the order of
$O_{{\cal D}'}({\varepsilon ^2})$ in ${\cal D}'$ we obtain

\begin{equation}
\label{56}
\begin{array}{rcl}
\displaystyle
[u^*(x,t,\varepsilon )]^2&=&u_0^2 + g^2\omega^2 + 2u_0g\omega +
2u_0e{\varepsilon }\omega_0^{-}, \\
\displaystyle
[u^*(x,t,\varepsilon )]^3&=&u_0^3+g^3\omega^3+3u_0g^2\omega^2+3u_0^2g\omega
+ 3u_0^2e{\varepsilon }\omega_0^{-}.
\end{array}
\end{equation}

    Then for all $\varphi(x) \in {\cal D}$:
$$
\int\omega\Big(\alpha(t)\frac{x-\phi(t)}{\varepsilon }\Big)\varphi(x)\,dx
= \frac{{\varepsilon }}{\alpha(t)}\int\omega(\eta)
\varphi\Big(\phi(t)+\frac{\varepsilon }{\alpha(t)}\eta\Big)\,d\eta
= \frac{\Omega_1}{\alpha(t)}{\varepsilon }\varphi(\phi(t))
+ O({\varepsilon ^2}),
$$
where $\Omega_1=\int\omega(\eta)\,d\eta= \int\ch^{-2}(\eta)\,d\eta= 2$.
Therefore,
$$
\omega\Big(\alpha(t)\frac{x-\phi(t)}{\varepsilon }\Big)
= \frac{\Omega_1}{\alpha(t)}{\varepsilon }\delta(x-\phi(t))
+ O_{{\cal D}'}({\varepsilon ^2}).
$$

   In a similar way we find that
$$
\omega^2\Big(\alpha(t)\frac{x-\phi(t)}{\varepsilon }\Big)
= \frac{\Omega_2}{\alpha(t)}{\varepsilon }\delta(x-\phi(t))
+ O_{{\cal D}'}({\varepsilon ^2}),
$$
$$
\omega^3\Big(\alpha(t)\frac{x-\phi(t)}{\varepsilon }\Big)
= \frac{\Omega_3}{\alpha(t)}{\varepsilon }\delta(x-\phi(t))
+ O_{{\cal D}'}({\varepsilon ^2}),
$$
where $\Omega_2= \int\omega^2(\eta)\,d\eta= \int\ch^{-4}(\eta)\,d\eta
=\frac{4}{3}$, \
$\Omega_3=\int\omega^3(\eta)\,d\eta= \int\ch^{-6}(\eta)\,d\eta
=\frac{16}{15}$.

    Now we find the asymptotics of the dispersion term
${\cal G}(u^*(x,t,\varepsilon ),\varepsilon )$. It immediately follows
from (\ref{52}) that
$$
{\cal G}(u^*(x,t,\varepsilon ),\varepsilon )
= {\varepsilon }^2[((u^*)^2)_{xx}-3(u^*_x)^2]_x
= -3{\varepsilon }^2[(u^*_x)^2]_x + O_{{\cal D}'}({\varepsilon ^2}).
$$
>From (\ref{54}) we have
$$
u^*_x(x,t,\varepsilon )= u_{0x}+g\omega_x + e_x{\varepsilon }\omega_0^{-}
+e{\varepsilon }\omega_{0x}^{-}.
$$

   Squaring this expression and analyzing its components, we see that up to
the terms of the order of $O_{{\cal D}'}({\varepsilon ^2})$ we have
$$
{\cal G}(u^*(x,t,\varepsilon ),\varepsilon )
= -3{\varepsilon }^2g^2[(\omega_x)^2]_x + O_{{\cal D}'}({\varepsilon ^2}).
$$

    After calculating the weak asymptotics of this expression, we obtain for
all $\varphi(x) \in {\cal D}$
$$
\langle{\cal G}(u^*(x,t,\varepsilon ),\varepsilon ),\varphi(\xi)\rangle
= -3g^2(t){\varepsilon }^2\int\bigg[
\Big(\omega\Big(\alpha(t)\frac{x-\phi(t)}{\varepsilon }\Big)\Big)_x\bigg]^2
\varphi(x)\,dx
$$
$$
= -3g^2(t){\varepsilon }\alpha(t)\int
[\omega'(\eta)]^2\varphi(\phi(t)+\frac{\varepsilon }{\alpha(t)}\eta)\,d\eta
= -3\Omega_4g^2(t)\alpha(t){\varepsilon }\varphi(\phi(t))
+ O({\varepsilon ^2}),
$$
where $\Omega_4= \int[\omega'(\eta)]^2\,d\eta
= 4\int\ch^{-6}(\eta)\sh^{2}(\eta)\,d\eta= \frac{16}{15}$.

   It follows that the weak asymptotics of the dispersion term is
\begin{equation}
\label{57}
{\cal G}(u^*_{\varepsilon }(x,t,),\varepsilon )
= -3g^2(t)\alpha(t)\Omega_4{\varepsilon }\delta'(x-\phi(t))
+ O_{{\cal D}'}({\varepsilon ^2}).
\end{equation}

    Substituting the asymptotics for $\omega$, \ $\omega^2$, \ $\omega^3$
obtained above into (\ref{54})--(\ref{56}) we have
$$
u^*_{\varepsilon }(x,t,)=u_0
+ g\frac{\Omega_1}{\alpha}{\varepsilon }\delta(x-\phi(t))
+ e(x,t){\varepsilon }\theta(-x+\phi(t)) + O_{{\cal D}'}({\varepsilon ^2}),
$$
$$
[u^*_{\varepsilon }(x,t)]^2 = u_0^2
+ \Big\{g^2\frac{\Omega_2}{\alpha} + 2u_0g\frac{\Omega_1}{\alpha}\Big\}
{\varepsilon }\delta(x-\phi(t))
+ 2u_0e{\varepsilon }\theta(-x+\phi(t)) + O_{{\cal D}'}({\varepsilon ^2}),
$$
$$
[u^*_{\varepsilon }(x,t)]^3=u_0^3
+ \Big\{g^3\frac{\Omega_3}{\alpha}+ 3u_0g^2\frac{\Omega_2}{\alpha}
+ 3u_0^2g\frac{\Omega_1}{\alpha}\Big\}{\varepsilon }\delta(x-\phi(t))
+ 3u_0^2e{\varepsilon }\theta(-x+\phi(t)) + O_{{\cal D}'}({\varepsilon ^2}).
$$

    After the substitution of the coefficients $\Omega_k$, \ $k= 1,2,3$,
into these expressions we obtain
\begin{equation}
\label{58}
\begin{array}{rcl}
\displaystyle
u^*_{\varepsilon }(x,t) = u_0(x,t)
+2\sqrt{6}g^{1/2}(t){\varepsilon }\delta(x-\phi(t))
\qquad\qquad\qquad\qquad\qquad\qquad &&\\
\displaystyle
+e(x,t){\varepsilon }\theta(-x+\phi(t))+O_{{\cal D}'}({\varepsilon ^2}),&&\\
\displaystyle
[u^*_{\varepsilon }(x,t,)]^2 = u_0^2(x,t)
+ \sqrt{6}\Big\{ \frac{4}{3}g^{3/2}(t) + 4u_0(\phi(t),t)g^{1/2}(t)\Big\}
{\varepsilon }\delta(-x+\phi(t)) &&\\
\displaystyle
+ 2u_0(x,t)e(x,t){\varepsilon }\theta(-x+\phi(t))
+ O_{{\cal D}'}({\varepsilon ^2}), &&\\
\displaystyle
[u^*_{\varepsilon }(x,t)]^3 = u_0^3(x,t)
+\sqrt{6}\Big\{\frac{16}{15}g^{5/2}(t) + 4u_0(\phi(t),t)g^{3/2}(t)
\qquad\qquad\qquad &&\\
\displaystyle
+ 6u_0^2(\phi(t),t)g^{1/2}(t)(\phi(t),t)\Big\}
{\varepsilon }\delta(x-\phi(t)) \qquad\qquad &&\\
\displaystyle
+ 3u_0^2(x,t)e(x,t){\varepsilon }\theta(-x+\phi(t))
+ O_{{\cal D}'}({\varepsilon ^2}). &&\\
\end{array}
\end{equation}

    According to Definition~\ref{de1}, substituting asymptotics (\ref{58})
into the first equation of system (\ref{55}) and setting the coefficients
of ${\varepsilon }^0$, \ ${\varepsilon }\delta$, \
${\varepsilon }\delta'$ and ${\varepsilon }\theta$ equal to zero, we
obtain the necessary and sufficient conditions for the right-hand side of
the equation $L_{KdV}[u]=0$ to be of order $O_{{\cal D}'}(\varepsilon ^2)$:
\begin{equation}
\label{59}
\begin{array}{rcrcl}
\displaystyle
\varepsilon ^0:\,& & u_{0t}+(u^2_0)_x &=&0,\\
\displaystyle
\varepsilon \delta':\,& & -\phi_{t}+2u_0(\phi(t),t)+\frac{2}{3}g(t)&=&0, \\
\displaystyle
\varepsilon \delta:\,& & \sqrt{6}\frac{g_t(t)}{g^{1/2}(t)}
- (-\phi_{t}+2u_0(\phi(t),t))e(\phi(t),t) &=&0,\\
\displaystyle
\varepsilon \theta:\,& & \big(e_{t}(x,t)
+ 2(u_0(x,t)e(x,t))_x \big)\Big|_{x < \phi(t)} &=&0.
\end{array}
\end{equation}

    The second and third equations of system (\ref{59}) imply the equation
for the jump of the amplitude of a small shock wave:
\begin{equation}
\label{60}
e(\phi(t),t)= -\frac{3}{2}\sqrt{6}\frac{g_t(t)}{g^{3/2}(t)}.
\end{equation}

    Following Definition~\ref{de1}, we now substitute asymptotics (\ref{58})
into the second equation of system (\ref{55}) and, setting the coefficients
of ${\varepsilon }^0$, \ ${\varepsilon }\delta$, \ ${\varepsilon }\delta'$
and ${\varepsilon }\theta$ equal to zero, we obtain the necessary and
sufficient conditions for the right-hand side of the equation
$uL_{KdV}[u]=0$ to be of order $O_{{\cal D}'}(\varepsilon ^2)$:
\begin{equation}
\label{61}
\begin{array}{rcrcl}
\displaystyle
\varepsilon ^0:\,&& (u_0^2)_t+\frac{4}{3}(u^3_0)_x &=&0,\\
\displaystyle
\varepsilon \delta':\,&& \Big(-\phi_{t}+2u_0(\phi(t),t)+\frac{2}{3}g(t)\Big)
\Big(g(t)+3u_0(\phi(t),t)\Big) &=&0, \\
\displaystyle
\varepsilon \delta:\,&& \sqrt{6}g(t)\Big(g(t)+2u_0(\phi(t),t)\Big)_t
\qquad\qquad\qquad\qquad\qquad\qquad\qquad &&\\
\displaystyle
\,&& -\frac{2}{3}u_0(\phi(t),t)g^{3/2}(t)
\Big(-e(\phi(t),t)-\frac{3}{2}\sqrt{6}\frac{g_t(t)}{g^{3/2}(t)}\Big)&=&0,\\
\displaystyle
\varepsilon \theta:\,&& \Big(\big(2u_0(x,t)e(x,t)\big)_{t}
+4\big(u_0^2(x,t)e(x,t)\big)_x \Big)\Big|_{x < \phi(t)} &=&0.
\end{array}
\end{equation}

    Since $u_0(x,t)$ and $e(x,t)$ are smooth functions, the first and
the last equations from systems (\ref{59}) and (\ref{61}) are equivalent.
It follows from the third equation of system (\ref{61}) and equation
(\ref{60}) that $g(t)+2u_0(\phi(t),t)={\mbox const}$.

    Relations (\ref{59}) -- (\ref{61}) imply the following theorem which
determines the dynamics of a single deformed soliton to the KdV equation.

\begin{th}
\label{th3}
   Let us assume that for $t \in [0, \, T]$ there exists a
smooth solution
$u_0(x,t)$ to the Hopf equation with the smooth initial condition
$u_0(x,t)\Big|_{t=0}=u_0^0(x)$.

   Then the KdV equation on the closed interval $[0, \ T]$, up to
$O_{{\cal D}'}(\varepsilon ^2)$,
has an infinitely narrow $\delta$-soliton-type solution
$$
u^*_{\varepsilon }(x,t) = u_0(x,t)
+2\sqrt{6}g^{1/2}(t){\varepsilon }\delta(x-\phi(t))
+e(x,t){\varepsilon }\theta(-x+\phi(t)),
$$
if and only if the unknown smooth functions $u_0(x,t)$, \ $g(t)$,
\ $e(x,t)$, \ $\phi(t)$ satisfy the system of equations
\begin{equation}
\label{62}
\begin{array}{rcl}
\displaystyle
u_{0t}+(u^2_0)_x &=&0,\\
\displaystyle
\phi_{t} &=& 2u_0(\phi(t),t)+\frac{2}{3}g(t), \\
\displaystyle
g(t)+2u_0(\phi(t),t) &=& g(0)+2u_0^0(\phi(0)) , \\
\displaystyle
		&&\\
\displaystyle
\big(e_{t}(x,t)+ 2(u_0(x,t)e(x,t))_x \big)\Big|_{x < \phi(t)} &=&0, \\
\displaystyle
e(\phi(t),t) &=& -\frac{3}{2}\sqrt{6}\frac{g_t(t)}{g^{3/2}(t)}. \\
\end{array}
\end{equation}
\end{th}

    Thus system of equations coinsides with system~(\ref{15})
studied in the Introduction.

\section{Infinitely narrow solitons to the equations $u_t+(f(u))_x= 0$
and $u_t+(f(u))_x + {\varepsilon }^2u_{xxx}= 0$ \/}%4
\label{s4}

    1. {\it The equation $u_t+(f(u))_x=0$.} Consider the infinitely
narrow soliton-type solution of this equation, substituting smooth ansatz
(\ref{24}) into the equation. This ansatz is a generalization of the
smooth one (\ref{50}) and has, up to $O_{{\cal D}'}({\varepsilon }^2)$,
a weak asymptotics of the form (\ref{25}). In this case, in (\ref{24})
$\omega_0(\frac{x}{\varepsilon })$ is an approximation of the Heaviside
function $\theta(x)$.

    By analogy to (\ref{35}), define the asymptotic distribution
$$
f\Big(u_0(x,t) + \Omega_1(t){\varepsilon }\delta(x)
+ e(x,t){\varepsilon }\theta(x)\Big)
$$
as the weak asymptotics of the approximating function as $\varepsilon \to +0$
$$
f\Big(u_0(x,t)+\omega\big(\frac{x}{\varepsilon },t\big)
+e(x,t){\varepsilon }\omega_0(\frac{x}{\varepsilon })\Big).
$$

    It is clear that
$$
f\Big(u_0+\omega\big(\frac{x}{\varepsilon },t\big)
+e{\varepsilon }\omega_0(\frac{x}{\varepsilon })\Big)
= f\Big(u_0+\omega(\frac{x}{\varepsilon },t)\Big)
+ {\varepsilon }f'(u_0)e\omega_0(\frac{x}{\varepsilon })
+O({\varepsilon }^2).
$$
In addition, according to the Lagrange theorem, we have
$$
f\Big(u_0+\omega(\frac{x}{\varepsilon },t)\Big)-f(u_0)
= f'\Big(u_0+\Theta \omega(\frac{x}{\varepsilon },t)\Big)
\omega(\frac{x}{\varepsilon },t),
$$
where $0<\Theta<1$. Hence, for the function
$f\Big(u_0+\omega(\frac{x}{\varepsilon },t)\Big)-f(u_0)$
we have the same estimation with respect to $\tau$
as for the function $\omega(\tau,t)$,

    Applying the function $f\Big(u_0+\omega(\frac{x}{\varepsilon },t)
+e{\varepsilon }\omega_0(\frac{x}{\varepsilon })\Big)$ to a test function
and performing the change of variables $x= {\varepsilon }\tau$, we obtain,
as in deduction of formula (\ref{35}),
$$
f\Big(u_0(x,t) + \omega(\frac{x}{\varepsilon },t)
+e{\varepsilon }\omega_0(\frac{x}{\varepsilon }\Big)
$$
\begin{equation}
\label{69}
= f(u_0(x,t)) + \Lambda(0,t){\varepsilon }\delta(x)
+ f'(u_0(x,t))e(x,t){\varepsilon }\theta(x)
+ O_{{\cal D}'}({\varepsilon }^2), \quad \varepsilon  \to +0,
\end{equation}
where the function
\begin{equation}
\label{70}
\Lambda(x,t)
=\int\Big[f\big(u_0(x,t)+\omega(\tau,t)\big)-f(u_0(x,t))\Big]\,d\tau.
\end{equation}

    Substituting $u^*_{\varepsilon }(x,t)$, in the form of singular
ansatz (\ref{25}), into the initial equation and using (\ref{69}) we obtain
$$
u^*_{{\varepsilon }t}+(f(u^*_{\varepsilon }))_x = u_{0t}+(f(u_0))_x
+\Big\{-\phi_{t}(t)\Omega_1(t) + \Lambda(\phi(t),t)\Big\}
{\varepsilon }\delta'(x-\phi(t))
$$
$$
+\Big\{(\Omega_1(t))_t - e(x,t)\phi_{t}(t) + e(x,t)f'(u_0(x,t))\Big\}
{\varepsilon }\delta(x-\phi(t))
$$
$$
+ \Big\{e_{t}(x,t) + \big(f'(u_0(x,t))e(x,t)\big)_x \Big\}
{\varepsilon }\theta(x-\phi(t)) + O_{{\cal D}'}({\varepsilon }^2),
\quad {\varepsilon } \to +0.
$$

    Using formula (\ref{48}) and setting the coefficients of
${\varepsilon }^0$, \ ${\varepsilon }\delta(x-\phi)$, \
${\varepsilon }\delta'(x-\phi)$, \ ${\varepsilon }\theta(x-\phi)$,
equal to zero, we arrive at the following result.

\begin{th}
\label{th4}
    Suppose that for $t \in [0, \ T]$ there exists a smooth
solution $u_0(x,t)$ to equation
$u_{0t}+(f(u_{0}))_x= 0$ with the smooth initial condition
$u_0(x,t)\Big|_{t=0}= u_0^0(x)$, where $f(u)$ is a smooth
function.

    Then, for $t \in [0, \ T]$, the initial equation  has the infinitely
narrow $\delta$-soliton solution {\rm (\ref{25})} if and only if the
unknown smooth functions $u_0(x,t)$, \ $e(x,t)$, \ $\phi(t)$ satisfy the
following system of equations:
\begin{equation}
\label{71}
\begin{array}{rcl}
\displaystyle
u_{0t}+(f(u_0))_x &=&0,\\
\displaystyle
-\phi_{t}\Omega_1(t) + \Lambda(\phi(t),t) &=&0, \\
\displaystyle
\big(\Omega_1\big)_t + \Big[f'(u_0(\phi(t),t))
-\frac{\Lambda(\phi(t),t)}{\Omega_1(t)}\Big]e(\phi(t),t) &=&0,\\
\displaystyle
\Big[e_{t}(x,t)+\big(f'(u_0(x,t))e(x,t)\big)_x\Big]\Big|_{x>\phi(t)}&=&0,
\end{array}
\end{equation}
where $\Omega_1(t)= \int\omega(\tau,t)\,d\tau$ and the function
$\Lambda(x,t)$ is defined in {\rm (\ref{70})}.
\end{th}

   It is easy to verify that if the inequality $f''(u)>0$ is satisfied,
then for $\Omega_1(t)>0$ system~(4.54) has exactly the same properties as
system~(\ref{14}).

    2. {\it The equation $u_t+(f(u))_x + {\varepsilon }^2u_{xxx}= 0$.}

    As it was said in Introduction, for some convex smooth $f(u)$ this
equation has exact one-soliton solutions. In this case we use our
approach to describe the dynamics of propagation of a deformed infinitely
narrow soliton. By virtue of Definition~\ref{de1}, as in the case of the
KdV equation, the infinitely narrow soliton-type solution (\ref{25}) is
defined as a solution of the two equations
\begin{equation}
\label{72}
\begin{array}{rcl}
\displaystyle
L[u]= u_t + (f(u))_x + {\varepsilon }^2u_{xxx} &=&
O_{{\cal D}'}(\varepsilon^2), \\
\displaystyle
uL[u]= (u^2)_t + ({\tilde f}(u))_x + {\varepsilon }^2 2uu_{xxx} &=&
O_{{\cal D}'}(\varepsilon^2), \\
\end{array}
\end{equation}
where  ${\tilde f}(u)= 2uf(u)-2\int_{0}^{u} f(z)\,dz$.
Note that the ansatz (\ref{25}) has the approximation (\ref{24}), where
$\omega_0= \omega_0^{-}(\frac{x}{\varepsilon })$ is the approximation
of the Heaviside function $\theta(-x)$.

    The dispersion term in the second conservation law is just the same
as for the KdV equation:
$$
{\cal G}(u,\varepsilon )={\varepsilon }^2 2uu_{xxx}
= {\varepsilon }^2[(u^2)_{xx}-3(u_x)^2]_x,
$$
and its weak asymptotics has the form (\ref{57}) and
$\Omega_{4}(t)= \int[\omega'_\tau(\tau,t)]^2\,d\tau$.

    Substituting the smooth ansatz (\ref{24}) into the initial
equation, just as for the KdV equation from Section~\ref{s3}.2,
we obtain,
$$
[u^*_{\varepsilon }(x,t)]^2 = u_0^2
+ \Big\{\Omega_2(t) + 2u_0\Omega_1(t)\Big\}{\varepsilon }\delta(x-\phi(t))
+ 2u_0e{\varepsilon }\theta(-x+\phi(t)) + O_{{\cal D}'}({\varepsilon }^2),
$$
where $\Omega_1(t)= \int\omega(\tau,t)\,d\tau$, \
$\Omega_2(t)= \int\omega^2(\tau,t)\,d\tau$.

   From (\ref{24}) and (\ref{25}), using formula (\ref{69}), we find
$$
{\tilde f}\big(u_0(x,t) + \Omega_1(t){\varepsilon }\delta(x)
+ e(x,t){\varepsilon }\theta(-x)\big)
$$
\begin{equation}
\label{73}
= {\tilde f}(u_0(x,t))
+{\tilde \Lambda}(0,t){\varepsilon }\delta(x)
+ {\tilde f}'(u_0(x,t))e(x,t){\varepsilon }\theta(-x)
+ O_{{\cal D}'}({\varepsilon }), \quad \varepsilon  \to +0,
\end{equation}
where ${\tilde f}'(u)= 2uf'(u)$ and
\begin{equation}
\label{74}
{\tilde \Lambda}(x,t)=\int\Big[{\tilde f}\big(u_0(x,t)+\omega(\tau,t)\big)
- {\tilde f}(u_0(x,t))\Big]\,d\tau.
\end{equation}

    Obviously, the system of equations derived from the first equation
in (\ref{72}) coincides with system (\ref{71}) from Theorem~\ref{th4}
which describes the dynamics of the formal soliton solution of the
equation $u_t+(f(u))_x= 0$.

    We find the system of equations which follows from the second
conservation law $uL[u]= 0$ in (\ref{72}). To this end, substitute the
asymptotics $[u^*_{\varepsilon }(x,t)]^2$, (\ref{73}) and (\ref{57})
into the second equation (\ref{72}) and set the coefficients of
${\varepsilon }^0$, \ ${\varepsilon }\delta$, \ ${\varepsilon }\delta'$
and ${\varepsilon }\theta$ equal to zero.

    Setting the coefficients of ${\varepsilon }^0$ and ${\varepsilon }\theta$
in the obtained system equal to zero, we have the equations
$$
\begin{array}{rcl}
\displaystyle
(u_0^2)_t + ({\tilde f}(u_0))_x &=& 0, \\
\displaystyle
\Big[(2u_0e)_{t} + \big(2u_0f'(u_0)e\big)_x\Big]\Big|_{x<\phi(t)} &=&0, \\
\end{array}
$$
which, in virtue of smoothness of the functions $u_0(x,t)$, \ $e(x,t)$,
\ $f(u)$, coincide with the corresponding equations of system (\ref{71})
of the first conservative law $L[u]=0$.

    After setting the coefficient of ${\varepsilon }\delta$ equal to zero,
we find
$$
\Big(\Omega_2(t) + 2u_0(\phi(t),t)\Omega_1(t)\Big)_t
+ 2u_0(\phi(t),t)e(\phi(t),t)\Big(f'(u_0) - \phi_{t}(t)\Big)=0.
$$

    Using the second and the third equations of system (\ref{71}), we
bring the last equation to the form
\begin{equation}
\label{75}
\frac{d}{d t}\Omega_2(t)+ 2\Omega_1(t)\frac{d}{d t}u_0(\phi(t),t)=0,
\end{equation}
which coincides with the third equation from (\ref{62}) for the KdV
equation when $\alpha(t)=\sqrt{\frac{g(t)}{6}}$.

    Consider the coefficient of ${\varepsilon }\delta'$ and prove
\begin{lem}
\label{lem1}
    The coefficients of ${\varepsilon }\delta'$ in the conservation laws
$L[u]=0$ and $uL[u]=0$ (\ref{72}) coincide.
\end{lem}

    {\it Proof.} Setting the coefficient of ${\varepsilon }\delta'$ in
the left-hand side of the equation $uL[u]=0$ equal to zero, we find:
\begin{equation}
\label{76}
-\phi_{t}(t)\Big(\Omega_2(t)+ 2u_0(\phi(t),t)\Omega_1(t)\Big) +
{\tilde \Lambda}(\phi(t),t)-3\Omega_{\Delta}(t)=0.
\end{equation}

    Let $u(x,t)=u_0(x,t) + \omega(\tau,t)$ be an exact soliton
solution to the equation $u_t+(f(u))_x + {\varepsilon }^2u_{xxx}=0$, where
$\tau=\frac{x-\phi(t)}{\varepsilon }$. Substitute $u(x,t)$ into this
equation and consider the coefficient of $\frac{1}{\varepsilon }$:
\begin{equation}
\label{77}
-\phi_t\omega_{\tau} + (f(u_0 + \omega))_{\tau} + \omega_{\tau\tau\tau}=0.
\end{equation}

    Integrating equation (\ref{77}) with respect to $\tau$ from
$-\infty$ to $\tau$ and taking into account that $\omega(\tau,t)$ and its
derivatives with respect to $\tau$ tend to zero as $|\tau| \to \infty$,
we obtain
\begin{equation}
\label{78}
-\phi_t\omega + f\big(u_0 + \omega\big) + \omega_{\tau\tau}= f(u_0).
\end{equation}
Here the integration constant $f(u_0)$ has been derived from the
boundary conditions.

    Multiplying equation (\ref{78}) by $\omega_{\tau}$ and integrating
it again with respect to $\tau$ from $-\infty$ to $\tau$, we have
\begin{equation}
\label{79}
-\phi_t\omega^2 +
2\int_{-\infty}^{\tau}f\big(u_0 + \omega\big)\omega_{\tau}\,d\tau
-2f(u_0)\int_{-\infty}^{\tau}\omega_{\tau}\,d\tau + (\omega_{\tau})^2= 0.
\end{equation}

    Let us introduce the function ${\tilde f}_1(u)=\int_{0}^{u}f(z)\,dz$, \
${\tilde f}_1(0)=0$.

    Since
$$
\frac{d}{d\tau}{\tilde f}_1\big(u_0 + \omega\big)
= f\big(u_0 + \omega\big)\omega_{\tau},
$$
and since $\omega(\tau,t)\to 0$ for $\tau \to -\infty$,
$$
\int_{-\infty}^{\tau}f\big(u_0 + \omega\big)\omega_{\tau}\,d\tau
=\int_{-\infty}^{\tau}\frac{d}{d\tau}{\tilde f}_1\big(u_0+\omega\big)\,d\tau
={\tilde f}_1\big(u_0 + \omega\big) - {\tilde f}_1\big(u_0\big)
$$
    Now equation (\ref{79}) can be rewritten in the form
$$
-\phi_t\omega^2
+ 2\Big[{\tilde f}_1\big(u_0 + \omega\big)- {\tilde f}_1\big(u_0\big)\Big]
- 2f(u_0)\omega + (\omega_{\tau})^2=0.
$$

    Integrating this equation with respect to $\tau$ from $-\infty$ to
$\infty$, we find
\begin{equation}
\label{80}
-\phi_t\Omega_2 -2f(u_0)\Omega_1 + 2{\tilde \Lambda}_1 + \Omega_{\Delta}=0,
\end{equation}
where $\Omega_1(t)=\int\omega(\tau,t)\,d\tau$, \
$\Omega_2(t)=\int\omega^2(\tau,t)\,d\tau$, \
$\Omega_{\Delta}(t)=\int(\omega_{\tau})^2(\tau,t)\,d\tau$,
\begin{equation}
\label{81}
{\tilde \Lambda}_1(x,t)
=\int\Big[{\tilde f}_1\big(u_0(x,t)+\omega(\tau,t)\big)
-{\tilde f}_1(u_0(x,t))\Big]\,d\tau.
\end{equation}

    Multiplying (\ref{78}) by $\omega$ and integrating the obtained
equation with respect to $\tau$ from $-\infty$ to $\infty$, we find
\begin{equation}
\label{82}
-\phi_t\Omega_2
+ \int\Big[f\big(u_0 + \omega\big)-f\big(u_0\big)\Big]\omega\,d\tau
- \Omega_{\Delta}=0,
\end{equation}
where the last term has been derived by integrating the expression
$\int\omega_{\tau\tau}\omega\,d\tau$ by parts.

    Subtracting equation (\ref{80}) from the doubled equation (\ref{82}),
we have
\begin{equation}
\label{83}
-\phi_t\Omega_2 + 2f(u_0)\Omega_1
+2\int\Big[f\big(u_0+\omega\big)-f\big(u_0\big)\Big]\omega\,d\tau
- 2{\tilde \Lambda}_1 - 3\Omega_{\Delta}=0.
\end{equation}

    Since ${\tilde f}(u)=2uf(u)-2{\tilde f}_1(u)$, where
${\tilde f}_1(u)=\int_{0}^{u} f(z)\,dz$, the function ${\tilde \Lambda}(x,t)$
from (\ref{74}) can be rewritten in the form
$$
{\tilde \Lambda}(x,t)
= 2\int\Big[(u_0+\omega)f(u_0+\omega) - u_0f(u_0)\Big]\,d\tau
-2\int\Big[{\tilde f}(u_0+\omega) - {\tilde f}(u_0)\Big]\,d\tau.
$$

    Adding and subtracting $f(u_0)\omega$ in the first integral and
comparing the obtained expression to formulae (\ref{69}) and (\ref{81}) for
$\Lambda(x,t)$ and ${\tilde \Lambda}_1(x,t)$, respectively, we obtain
\begin{equation}
\label{84}
{\tilde \Lambda}(x,t)=2u_0\Lambda(x,t)+2f(u_0)\Omega_1
+2\int\Big[f(u_0+\omega)-f(u_0)\Big]\omega\,d\tau -2{\tilde \Lambda}_1(x,t).
\end{equation}

   Setting the coefficient of ${\varepsilon }\delta'$ in the first
conservation law $L[u]= 0$ equal to zero and taking into account
Theorem~\ref{th4}, we arrive at the equation
$$
-\phi_{t}\Omega_1(t) + \Lambda(\phi(t),t) = 0.
$$

    Substituting this relation and relation (\ref{84}) into equation
(\ref{83}), we obtain equation (\ref{76}), that is, the coefficient of
${\varepsilon }\delta'$ in the second conservation law $uL[u]=0$.
Thus, the equations obtained by setting the coefficients of
${\varepsilon }\delta'$ in both conservation laws (\ref{72}) equal to
zero are equivalent.

    The proof of the lemma is complete.

    As in Section~\ref{s3}.2, this implies the theorem which
determines the dynamics of propagation of the soliton solution
of equation under consideration.

\begin{th}
\label{th5}
    Suppose that for $t \in [0, \, T]$ there exists a smooth
solution $u_0(x,t)$ to the equation $u_t+(f(u))_x=0$ with the
smooth initial condition $u_0(x,t)\Big|_{t=0}=u_0^0(x)$.
Suppose that the problem~{\rm(1.21)} has a solution
and $f''(u)>0$ for $u\in[\omega_1,\omega_2]$.

    Then the equation $u_t+(f(u))_x + {\varepsilon }^2u_{xxx}=0$ for
$t \in [0, \ T]$ has an infinitely narrow $delta$-soliton-type solution
{\rm (\ref{25})} if and only if the unknown smooth functions $u_0(x,t)$,
\ $g(t)$, \ $e(x,t)$, \ $\phi(t)$ satisfy the following system of
equations:
\begin{equation}
\label{85}
\begin{array}{rcl}
\displaystyle
u_{0t}+(f(u_0))_x &=&0,\\
\displaystyle
-\phi_{t}\Omega_1(t) + \Lambda(\phi(t),t) &=&0, \\
\displaystyle
\frac{d}{d t}\Omega_2(t)+ 2\Omega_1(t)\frac{d}{d t}u_0(\phi(t),t)&=&0, \\
\displaystyle
\big(\Omega_1\big)_t + \Big[f'(u_0(\phi(t),t))
-\frac{\Lambda(\phi(t),t)}{\Omega_1(t)}\Big]e(\phi(t),t) &=&0,\\
\displaystyle
\Big[e_{t}(x,t)+\big(f'(u_0(x,t))e(x,t)\big)_x\Big]\Big|_{x<\phi(t)}&=&0, \\
\end{array}
\end{equation}
where $\Omega_1=\int\omega(\tau,t)\,d\tau$, \
$\Omega_2=\int\omega^2(\tau,t)\,d\tau$,
$$
\Lambda(\phi(t),t)=\int
\big[f\big(u_0(\phi(t),t)+g(t)\omega(\tau,t)\big)
-f(u_0(\phi(t),t))\big]\,d\tau.
$$
\end{th}

    It is easy to see that the system given in Theorem~\ref{th5} and
system~(\ref{15}) discussed in the Introduction are quite similar.

    In conclusion, we note that all constructions carred out in this
paper are formal and the weak asymptotics method is still not completely
developed.

    This work was partially supported by the Russian Foundation
for Basic Research (Grant Nos. 97-01-01123 and 99-01-01074).

\end{sloppypar}

\end {document}